\documentclass{Interspeech}

\interspeechcameraready

\usepackage{amsmath,graphicx}
\usepackage{amsfonts}
\usepackage{booktabs}
\usepackage[subrefformat=parens]{subcaption}
\usepackage{siunitx}
\usepackage{mathtools}
\usepackage{comment}
\usepackage{multirow}
\usepackage{makecell}
\usepackage{threeparttable}
\usepackage{arydshln}   
\usepackage[normalem]{ulem} 
\usepackage{cite}

\usepackage{tikz}
\usetikzlibrary{calc}
\usetikzlibrary{tikzmark}
\usetikzlibrary{arrows, positioning,intersections,angles}
\usetikzlibrary{fit}
\usetikzlibrary{decorations.pathreplacing}
\usetikzlibrary{decorations.pathmorphing}

\def\equationautorefname~#1\null{(#1\null)}

\renewcommand{\sectionautorefname}{Section}
\renewcommand{\subsectionautorefname}{\sectionautorefname}

\let\orgautoref\autoref

\renewcommand{\autoref}[1]
{%
\def\figureautorefname{Fig.}%
\def\subfigureautorefname{\figureautorefname}%
\def\sectionautorefname{Sec.}%
\def\subsectionautorefname{\sectionautorefname}%
\def\subsectionautorefname{\sectionautorefname}%
\orgautoref{#1}%
}

\newcommand{\vect}[1]{\mbox{\boldmath $#1$}}

\def\appendixautorefname~#1\null{~#1 \null}

\newcommand{\svss}{\textit{one-vs-one}}

\makeatletter
   {\list{}{\leftmargin=10pt 
            \labelwidth\z@ \itemindent-\leftmargin
            }}%
   {\endlist}
\makeatother

\newcommand{\customdashline}[1]{%
  \noalign{\vskip\aboverulesep} 
  \cdashline{#1}
  \noalign{\vskip\belowrulesep} 
}

\appto\TPTnoteSettings{\footnotesize}

\title{Pretraining Multi-Speaker Identification for Neural Speaker Diarization}

\author[]{Shota}{Horiguchi}
\author[]{Atsushi}{Ando}
\author[]{Marc}{Delcroix}
\author[]{Naohiro}{Tawara}

\affiliation[nocounter]{}{NTT Corporation}{Japan}

\email{horiguchi@ieee.org}

\begin{document}
\abovedisplayskip=1pt
\belowdisplayskip=1pt

\setlength\textfloatsep{5pt}
\setlength\dbltextfloatsep{5pt}
\setlength\floatsep{5pt}
\setlength\abovecaptionskip{2pt}
\setlength\belowcaptionskip{2pt}
\captionsetup[subfloat]{aboveskip=2pt,belowskip=4pt}

\maketitle
\begin{abstract}
End-to-end speaker diarization enables accurate overlap-aware diarization by jointly estimating multiple speakers' speech activities in parallel.
This approach is data-hungry, requiring a large amount of labeled conversational data, which cannot be fully obtained from real datasets alone.
To address this issue, large-scale simulated data is often used for pretraining, but it requires enormous storage and I/O capacity, and simulating data that closely resembles real conversations remains challenging.
In this paper, we propose pretraining a model to identify multiple speakers from an input fully overlapped mixture as an alternative to pretraining a diarization model.
This method eliminates the need to prepare a large-scale simulated dataset while leveraging large-scale speaker recognition datasets for training.
Through comprehensive experiments, we demonstrate that the proposed method enables a highly accurate yet lightweight local diarization model without simulated conversational data.
\end{abstract}

\section{Introduction}
Speaker diarization, which estimates who is speaking when, plays an essential role in multi-speaker applications such as speech separation~\cite{boeddeker2018front} and speech recognition~\cite{polok2024but,polok2024dicow}.
Speaker diarization methods fall into three directions: clustering speaker embeddings from short segments~\cite{landini2022bayesian}, end-to-end neural networks identifying speaker-wise speech activity~\cite{fujita2019end1}, and their hybrid methods~\cite{kinoshita2021integrating,horiguchi2021towards,bredin2023pyannote,plaquet2023powerset}.
The hybrid methods are especially promising because they can handle overlapping speech like end-to-end methods, while having the flexibility to handle unlimited speakers and long-form recordings like clustering-based methods.
This paper examines the improvement of the end-to-end model's performance using a hybrid approach.

To train end-to-end models, conversational recordings with speaker-wise activity labels are required.
Such data is limited by high annotation costs, so simulated mixtures generated from single-speaker utterances are commonly used for pretraining~\cite{fujita2019end1}.
However, this approach has several drawbacks.
First, simulated data is storage-unfriendly, typically requiring hundreds to thousands of gigabytes of storage.
While on-the-fly simulation can mitigate this~\cite{maiti2021endtoend}, it requires loading many utterances per mixture, leading to increased random access to the storage disk and significantly longer training times.
Second, model performance is highly sensitive to the quality of simulated data.
Early end-to-end neural diarization (EEND) suffered from dialogue act mismatches (e.g., turn-taking, backchannels) between simulated and real data, impairing pretraining effectiveness.
Making the dialog acts in simulated data more realistic improves performance~\cite{yamashita2022improving,landini2022from}, but per-domain simulation is infeasible due to storage limits.
Moreover, the optimality of mixture simulation methods remains uncertain, and investigating alternatives is equally challenging for the same reasons.

Another common practice is to perform pretraining using a compound of multiple real datasets~\cite{bredin2023pyannote,plaquet2023powerset,han2025leveraging}.
While it effectively expands the training dataset without concerns about simulation quality, its dynamics are difficult to predict.
Small variations in the combination can lead to significant differences in diarization performance (see the results in \cite{bredin2023pyannote} and \cite{plaquet2023powerset}).
Moreover, while compound datasets provide a potential solution, the amount is still significantly smaller than that of simulated data (see \autoref{tbl:dataset}), and whether they can achieve comparable accuracy has yet to be thoroughly investigated.

In this paper, we would like to answer the following research question: Can we build a powerful diarization system without relying on large simulated or real conversational data for pretraining?
To answer this question, we explore an alternative method, which relies on pretraining the encoder of a diarization model in a multi-speaker identification manner~\cite{horiguchi2024recursive}.
The evaluations on six datasets demonstrate that the proposed method not only alleviates storage and quality issues associated with simulated data and the scarcity of real data but also outperforms systems pretrained on these conversational data.
Note that using pretrained models based on self-supervised learning (SSL) for feature extraction in diarization has been explored before~\cite{cord2023frame,alvarez2024leveraging}, but they greatly increase the model size and complexity.
Our method also outperformed the SSL-based method with a significantly smaller number of parameters.

\section{Review of conventional methods}

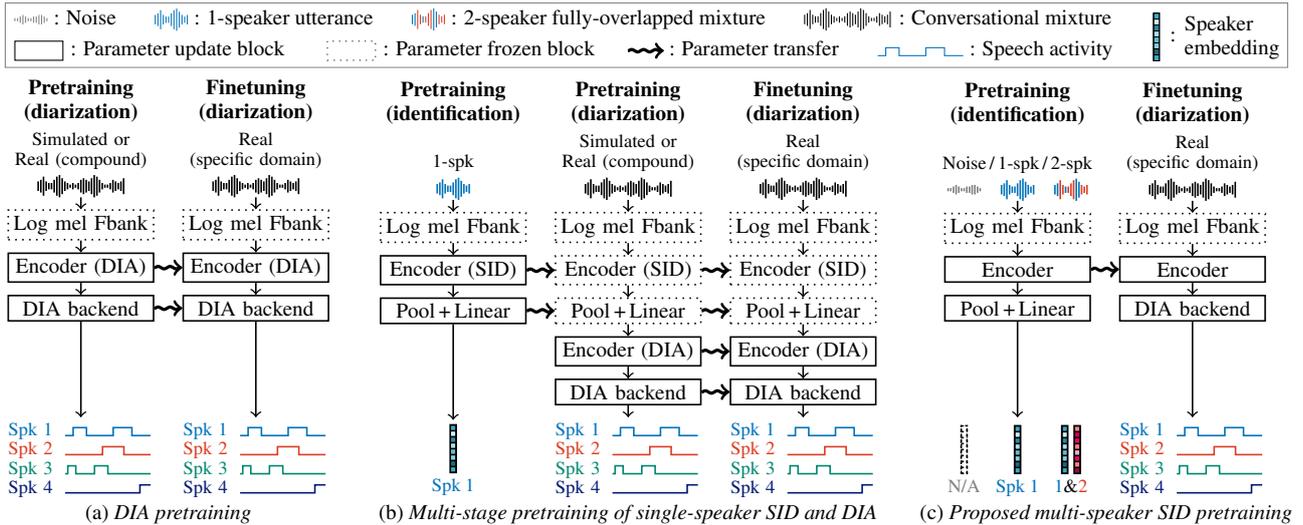
\begin{figure*}[t]
    \centering
    \begin{tikzpicture}
\definecolor{spkblue}{HTML}{0072BC}
\definecolor{spkred}{HTML}{DB3D23}
\definecolor{spkgreen}{HTML}{008770}
\definecolor{spkcyan}{HTML}{0FC8F2}
\definecolor{spkteal}{HTML}{2CD5B6}
\definecolor{spkdarkblue}{HTML}{001973}

\def\noise{{%
1,1,2,3,2,1,1,2,2,3,3,2,1,
}}
\def\utterance{{%
3,4,6,3,2,2,4,6,7,5,3,2,3,
}}
\def\conversation{{%
3,4,6,3,2,2,4,6,7,5,3,2,1,
2,1,2,4,5,4,3,5,7,6,4,2,
1,1,2,4,3,2,3,5,6,5,2,2,
}}

\newcommand{\drawNoiseWav}[3]{
    \path (#1); 
    \pgfgetlastxy{\startX}{\y}
    \begin{scope}[local bounding box=#3]
        \pgfmathsetmacro{\linespacing}{1.0}
        \foreach \i in {0,...,12} {
            \pgfmathsetmacro{\xx}{\startX + \i * \linespacing}
            \pgfmathsetmacro{\amp}{0.6 * \noise[\i]}%
            \draw[color=#2] (\xx pt, \y - \amp pt) -- (\xx pt, \y + \amp pt);
        }
    \end{scope}
}
\newcommand{\drawSingleWav}[3]{
    \path (#1); 
    \pgfgetlastxy{\startX}{\y}
    \begin{scope}[local bounding box=#3]
        \pgfmathsetmacro{\linespacing}{1.0}
        \foreach \i in {0,...,12} {
            \pgfmathsetmacro{\xx}{\startX + \i * \linespacing}
            \pgfmathsetmacro{\amp}{0.6 * \utterance[\i]}%
            \draw[color=#2] (\xx pt, \y - \amp pt) -- (\xx pt, \y + \amp pt);
        }
    \end{scope}
}
\newcommand{\drawMultiWav}[4]{    
    \path (#1); 
    \pgfgetlastxy{\startX}{\y}
    \begin{scope}[local bounding box=#4]
        \pgfmathsetmacro{\linespacing}{1.0}
        \foreach \i in {0,...,12} {
            \pgfmathsetmacro{\xx}{\startX + \i * \linespacing}
            \pgfmathsetmacro{\amp}{0.6 * \utterance[\i]}
            \pgfmathparse{int(\i==0||\i==1||\i==4||\i==8||\i==9)}
            \ifnum\pgfmathresult=1
                \draw[color=#2] (\xx pt, \y - \amp pt) -- (\xx pt, \y + \amp pt);
            \else
                \draw[color=#3] (\xx pt, \y - \amp pt) -- (\xx pt, \y + \amp pt);
            \fi
        }
    \end{scope}
}
\newcommand{\drawConversationWav}[3]{
    \path (#1); 
    \pgfgetlastxy{\startX}{\y}
    \begin{scope}[local bounding box=#3]
        \pgfmathsetmacro{\linespacing}{1.0}
        \foreach \i in {0,...,32} {
            \pgfmathsetmacro{\xx}{\startX + \i * \linespacing}
            \pgfmathsetmacro{\amp}{0.6 * \conversation[\i]}%
            \draw[color=#2] (\xx pt, \y - \amp pt) -- (\xx pt, \y + \amp pt);
        }
    \end{scope}
}

\newcommand{\coloredVector}[2]{%
    \foreach \y in {1,...,#1} {
        \pgfmathtruncatemacro\colorindex{#2[\y-1]}
        \fill[mycolor\colorindex] (0em,\y*0.25em-0.25em) rectangle (0.25em,\y*0.25em);
    }
}

\definecolor{mycolor1}{HTML}{FFD5E5}
\definecolor{mycolor2}{HTML}{FFAACC}
\definecolor{mycolor3}{HTML}{FF8082}
\definecolor{mycolor4}{HTML}{FF5599}
\definecolor{mycolor5}{HTML}{FF2A7F}
\definecolor{mycolor6}{HTML}{FF0066}
\definecolor{mycolor7}{HTML}{D40055}
\definecolor{mycolor8}{HTML}{AA0044}
\definecolor{mycolor9}{HTML}{D7EEF4}
\definecolor{mycolor10}{HTML}{AFDDE9}
\definecolor{mycolor11}{HTML}{87CDDE}
\definecolor{mycolor12}{HTML}{5FBCD3}
\definecolor{mycolor13}{HTML}{37ABC6}
\definecolor{mycolor14}{HTML}{2C89A0}
\definecolor{mycolor15}{HTML}{216776}
\definecolor{mycolor16}{HTML}{164450}

\def\embedfirst{{%
13,15,11,12,10,14,9,13,
}}
\def\embedsecond{{%
8,3,6,2,3,6,7,3,
}}

\drawNoiseWav{$(0.5em,-0.4em)$}{gray}{noise};
\node[draw=none,align=left,anchor=west,inner sep=0.1em] at ($(noise.west) + (1.5em,0.1em)$) (noise_label) {\footnotesize{: Noise}};

\drawSingleWav{$(noise_label.east) + (1.5em,-0.1em)$}{spkblue}{wav1};
\node[draw=none,align=left,anchor=base west,inner sep=0.1em] at ($(noise_label.base east) + (3em,0em)$) (1spk_label) {\footnotesize{: 1-speaker utterance}};

\drawMultiWav{$(1spk_label.east) + (1.5em,0em)$}{spkblue}{spkred}{wav2};
\node[draw=none,align=left,anchor=base west,inner sep=0.1em] at ($(1spk_label.base east) + (3em,0em)$) (2spk_label) {\footnotesize{: 2-speaker fully-overlapped mixture}};

\drawConversationWav{$(2spk_label.east) + (1.5em,0em)$}{black}{wav3};
\node[draw=none,align=left,anchor=base west,inner sep=0.1em] at ($(2spk_label.base east) + (5.3em,0em)$) (conv_label) {\footnotesize{: Conversational mixture}};

\draw (0.5em,-1.3em) rectangle (2.5em, -2.1em);
\node[draw=none,align=left,anchor=west,inner sep=0.1em] at (2.7em,-1.7em) (block1) {\footnotesize{: Parameter update block}};
\draw[dotted] (13.5em,-1.3em) rectangle (15.5em, -2.1em);
\node[draw=none,align=left,anchor=base west,inner sep=0.1em] at ($(block1.base west) + (13em,0em)$) (block2) {\footnotesize{: Parameter frozen block}};
\draw[->,very thick,decorate,decoration={snake,amplitude=0.05em,segment length=0.5em}] (26em,-1.7em) -- (27.5em,-1.7em);
\node[draw=none,align=left,anchor=base west,inner sep=0.1em] at ($(block2.base west) + (11.9em,0em)$) (transfer) {\footnotesize{: Parameter transfer}};
\draw[color=spkblue] ($(transfer.base east) + (1.5em,0pt)$) -- ++(3pt,0pt) -- ++(0pt,3pt) -- ++(5pt,0pt) -- ++(0pt,-3pt) -- ++(10pt,0pt) -- ++(0pt,3pt) -- ++(7pt,0pt) -- ++(0pt,-3pt) -- ++(7pt,0pt);
\node[draw=none,align=left,anchor=base west,inner sep=0.1em] at ($(transfer.base west) + (12.5em,0em)$) () {\footnotesize{: Speech activity}};

\node[rectangle,draw,text width=1*0.25em,
        text height=8*0.25em,inner sep=0,below right=-0.2em and 1.6em of conv_label.east,
        path picture={
            \coloredVector{8}{\embedfirst};
            \draw[step=0.25em, black] (path picture bounding box.south west) grid (path picture bounding box.north east);
        }
    ] (emb) {};
\node[draw=none,align=left,anchor=base west,inner sep=0.1em,right=0.3em of emb.east,font=\footnotesize] {:};
\node[draw=none,align=left,anchor=base west,inner sep=0.1em,right=0.9em of emb.east,font=\footnotesize] (emb_label) {Speaker\\embedding};

\draw[ultra thin,color=gray,name=legend] (0.1em,0.3em) rectangle (483.69684pt-0.3em, -2.4em);
\draw[draw=none,ultra thin,color=gray,name=legend] (0.1em,0.3em) rectangle (483.69684pt-0.3em, -2.4em-0.5mm);
\end{tikzpicture}\\
    \subfloat[DIA pretraining]{\begin{tikzpicture}[semithick,auto,
block/.style={
    rectangle,
    draw,
    text centered,
    text width=5.6em,
    inner sep=2pt,
    minimum height=1em,
    minimum width=5.6em,
    font=\footnotesize},
block_frozen/.style={
    rectangle,
    draw,
    dotted,
    text centered,
    text width=5.6em,
    inner sep=2pt,
    minimum height=1em,
    minimum width=5.6em,
    font=\footnotesize},
label/.style={
    draw=none,
    align=center,
    font=\small,
    inner sep=0,
    outer sep=0
},
]

\definecolor{spkblue}{HTML}{0072BC}
\definecolor{spkred}{HTML}{DB3D23}
\definecolor{spkgreen}{HTML}{008770}
\definecolor{spkcyan}{HTML}{0FC8F2}
\definecolor{spkteal}{HTML}{2CD5B6}
\definecolor{spkdarkblue}{HTML}{001973}

\def\utterance{{%
3,4,6,3,2,2,4,6,7,5,3,2,3,
}}
\def\conversation{{%
3,4,6,3,2,2,4,6,7,5,3,2,1,
2,1,2,4,5,4,3,5,7,6,4,2,
1,1,2,4,3,2,3,5,6,5,2,2,
}}

\newcommand{\drawNoiseWav}[3]{
    \path (#1); 
    \pgfgetlastxy{\startX}{\y}
    \begin{scope}[local bounding box=#3]
        \pgfmathsetmacro{\linespacing}{1.0}
        \foreach \i in {0,...,12} {
            \pgfmathsetmacro{\xx}{\startX + \i * \linespacing}
            \pgfmathsetmacro{\amp}{0.6 * \noise[\i]}%
            \draw[color=#2] (\xx pt, \y - \amp pt) -- (\xx pt, \y + \amp pt);
        }
    \end{scope}
}
\newcommand{\drawSingleWav}[3]{
    \path (#1); 
    \pgfgetlastxy{\startX}{\y}
    \begin{scope}[local bounding box=#3]
        \pgfmathsetmacro{\linespacing}{1.0}
        \foreach \i in {0,...,12} {
            \pgfmathsetmacro{\xx}{\startX + \i * \linespacing}
            \pgfmathsetmacro{\amp}{0.6 * \utterance[\i]}%
            \draw[color=#2] (\xx pt, \y - \amp pt) -- (\xx pt, \y + \amp pt);
        }
    \end{scope}
}
\newcommand{\drawMultiWav}[4]{    
    \path (#1); 
    \pgfgetlastxy{\startX}{\y}
    \begin{scope}[local bounding box=#3]
        \pgfmathsetmacro{\linespacing}{1.0}
        \foreach \i in {0,...,12} {
            \pgfmathsetmacro{\xx}{\startX + \i * \linespacing}
            \pgfmathsetmacro{\amp}{0.6 * \noise[\i]}
            \pgfmathparse{int(\i==0||\i==1||\i==4||\i==8||\i==9)}
            \ifnum\pgfmathresult=1
                \draw[color=#2] (\xx pt, \y - \amp pt) -- (\xx pt, \y + \amp pt);
            \else
                \draw[color=#3] (\xx pt, \y - \amp pt) -- (\xx pt, \y + \amp pt);
            \fi
        }
    \end{scope}
}
\newcommand{\drawConversationWav}[3]{
    \path (#1); 
    \pgfgetlastxy{\startX}{\y}
    \begin{scope}[local bounding box=#3]
        \pgfmathsetmacro{\linespacing}{1.0}
        \foreach \i in {0,...,32} {
            \pgfmathsetmacro{\xx}{\startX + \i * \linespacing}
            \pgfmathsetmacro{\amp}{0.6 * \conversation[\i]}%
            \draw[color=#2] (\xx pt, \y - \amp pt) -- (\xx pt, \y + \amp pt);
        }
    \end{scope}
}


\drawConversationWav{$(0pt,0pt)$}{black}{wav1};
\node[block,draw=none,text width=6em,font=\bfseries\footnotesize,above=2em of wav1] {Pretraining\\(diarization)};
\node[block,draw=none,text width=6em,font=\scriptsize,above=0em of wav1] {Simulated or\\Real (compound)};
\node[block_frozen,below=0.5em of wav1] (feat) {Log mel Fbank};
\node[block,below=0.5em of feat] (encoder) {Encoder (DIA)};
\node[block,below=0.5em of encoder] (diar) {DIA backend};

\node[anchor=north west,spkblue,text width=2em,inner sep=0,font=\scriptsize] at ($(diar.south west) + (0,-0.7em-31.6pt)$) (spk1_label) {Spk 1};
\draw[color=spkblue] ($(spk1_label.base east) + (3pt,0pt)$) -- ++(3pt,0pt) -- ++(0pt,3pt) -- ++(5pt,0pt) -- ++(0pt,-3pt) -- ++(10pt,0pt) -- ++(0pt,3pt) -- ++(7pt,0pt) -- ++(0pt,-3pt) -- ++(7pt,0pt);
\node[anchor=north west,spkred,text width=2em,inner sep=0,font=\scriptsize] at ($(spk1_label.north west) + (0,-0.8em)$) (spk2_label) {Spk 2};
\draw[color=spkred] ($(spk2_label.base east) + (3pt,0pt)$) -- ++(14pt,0pt) -- ++(0pt,3pt) -- ++(8pt,0pt) -- ++(0pt,-3pt) -- ++(10pt,0pt);
\node[anchor=north west,spkgreen,text width=2em,inner sep=0,font=\scriptsize] at ($(spk2_label.north west) + (0,-0.8em)$) (spk3_label) {Spk 3};
\draw[color=spkgreen] ($(spk3_label.base east) + (3pt,0pt)$) -- ++(1pt,0pt) -- ++(0pt,3pt) -- ++(3pt,0pt) -- ++(0pt,-3pt) -- ++(7pt,0pt) -- ++(0pt,3pt) -- ++(5pt,0pt) -- ++(0pt,-3pt) -- ++(16pt,0pt);
\node[anchor=north west,spkdarkblue,text width=2em,inner sep=0,font=\scriptsize] at ($(spk3_label.north west) + (0,-0.8em)$) (spk4_label) {Spk 4};
\draw[color=spkdarkblue] ($(spk4_label.base east) + (3pt,0pt)$) -- ++(28pt,0pt) -- ++(0pt,3pt) -- ++(4pt,0pt);

\drawConversationWav{$(wav1) + (5.5em,0em)$}{black}{wav2};
\node[block,draw=none,text width=6em,font=\bfseries\footnotesize,above=2em of wav2] {Finetuning\\(diarization)};
\node[block,draw=none,text width=5.5em,font=\scriptsize,above=0em of wav2] {Real\\(specific domain)};
\node[block_frozen,below=0.5em of wav2] (feat2) {Log mel Fbank};
\node[block,below=0.5em of feat2] (encoder2) {Encoder (DIA)};
\node[block,below=0.5em of encoder2] (diar2) {DIA backend};

\node[anchor=north west,spkblue,text width=2em,inner sep=0,font=\scriptsize] at ($(diar2.south west) + (0,-0.7em-31.6pt)$) (spk1_label) {Spk 1};
\draw[color=spkblue] ($(spk1_label.base east) + (3pt,0pt)$) -- ++(3pt,0pt) -- ++(0pt,3pt) -- ++(5pt,0pt) -- ++(0pt,-3pt) -- ++(10pt,0pt) -- ++(0pt,3pt) -- ++(7pt,0pt) -- ++(0pt,-3pt) -- ++(7pt,0pt);
\node[anchor=north west,spkred,text width=2em,inner sep=0,font=\scriptsize] at ($(spk1_label.north west) + (0,-0.8em)$) (spk2_label) {Spk 2};
\draw[color=spkred] ($(spk2_label.base east) + (3pt,0pt)$) -- ++(14pt,0pt) -- ++(0pt,3pt) -- ++(8pt,0pt) -- ++(0pt,-3pt) -- ++(10pt,0pt);
\node[anchor=north west,spkgreen,text width=2em,inner sep=0,font=\scriptsize] at ($(spk2_label.north west) + (0,-0.8em)$) (spk3_label) {Spk 3};
\draw[color=spkgreen] ($(spk3_label.base east) + (3pt,0pt)$) -- ++(1pt,0pt) -- ++(0pt,3pt) -- ++(3pt,0pt) -- ++(0pt,-3pt) -- ++(7pt,0pt) -- ++(0pt,3pt) -- ++(5pt,0pt) -- ++(0pt,-3pt) -- ++(16pt,0pt);
\node[anchor=north west,spkdarkblue,text width=2em,inner sep=0,font=\scriptsize] at ($(spk3_label.north west) + (0,-0.8em)$) (spk4_label) {Spk 4};
\draw[color=spkdarkblue] ($(spk4_label.base east) + (3pt,0pt)$) -- ++(28pt,0pt) -- ++(0pt,3pt) -- ++(4pt,0pt);

\draw[->] (wav1) -- (feat);
\draw[->] (feat) -- (encoder);
\draw[->] (encoder) -- (diar);
\draw[->] (diar) -- ($(diar.south) + (0,-0.5em-31.6pt)$);
\draw[->] (wav2) -- (feat2);
\draw[->] (feat2) -- (encoder2);
\draw[->] (encoder2) -- (diar2);
\draw[->] (diar2) -- ($(diar2.south) + (0,-0.5em-31.6pt)$);

\draw[->,very thick,decorate,decoration={snake,amplitude=0.05em,segment length=0.5em}] (encoder) -- (encoder2);
\draw[->,very thick,decorate,decoration={snake,amplitude=0.05em,segment length=0.5em}] (diar) -- (diar2);

\end{tikzpicture}\label{fig:conventional_eend}}
    \hfill
    \subfloat[Multi-stage pretraining of single-speaker SID and DIA]{\begin{tikzpicture}[semithick,auto,
block/.style={
    rectangle,
    draw,
    text centered,
    text width=5.6em,
    inner sep=2pt,
    minimum height=1em,
    minimum width=5.6em,
    font=\footnotesize},
block_frozen/.style={
    rectangle,
    draw,
    dotted,
    text centered,
    text width=5.6em,
    inner sep=2pt,
    minimum height=1em,
    minimum width=5.6em,
    font=\footnotesize},
label/.style={
    draw=none,
    align=center,
    font=\small,
    inner sep=0,
    outer sep=0
},
]

\definecolor{spkblue}{HTML}{0072BC}
\definecolor{spkred}{HTML}{DB3D23}
\definecolor{spkgreen}{HTML}{008770}
\definecolor{spkcyan}{HTML}{0FC8F2}
\definecolor{spkteal}{HTML}{2CD5B6}
\definecolor{spkdarkblue}{HTML}{001973}

\def\noise{{%
1,1,2,3,2,1,1,2,2,3,3,2,1,
}}
\def\utterance{{%
3,4,6,3,2,2,4,6,7,5,3,2,3,
}}
\def\conversation{{%
3,4,6,3,2,2,4,6,7,5,3,2,1,
2,1,2,4,5,4,3,5,7,6,4,2,
1,1,2,4,3,2,3,5,6,5,2,2,
}}

\newcommand{\drawNoiseWav}[3]{
    \path (#1); 
    \pgfgetlastxy{\startX}{\y}
    \begin{scope}[local bounding box=#3]
        \pgfmathsetmacro{\linespacing}{1.0}
        \foreach \i in {0,...,12} {
            \pgfmathsetmacro{\xx}{\startX + \i * \linespacing}
            \pgfmathsetmacro{\amp}{0.6 * \noise[\i]}%
            \draw[color=#2] (\xx pt, \y - \amp pt) -- (\xx pt, \y + \amp pt);
        }
    \end{scope}
}
\newcommand{\drawSingleWav}[3]{
    \path (#1); 
    \pgfgetlastxy{\startX}{\y}
    \begin{scope}[local bounding box=#3]
        \pgfmathsetmacro{\linespacing}{1.0}
        \foreach \i in {0,...,12} {
            \pgfmathsetmacro{\xx}{\startX + \i * \linespacing}
            \pgfmathsetmacro{\amp}{0.6 * \utterance[\i]}%
            \draw[color=#2] (\xx pt, \y - \amp pt) -- (\xx pt, \y + \amp pt);
        }
    \end{scope}
}
\newcommand{\drawMultiWav}[4]{    
    \path (#1); 
    \pgfgetlastxy{\startX}{\y}
    \begin{scope}[local bounding box=#4]
        \pgfmathsetmacro{\linespacing}{1.0}
        \foreach \i in {0,...,12} {
            \pgfmathsetmacro{\xx}{\startX + \i * \linespacing}
            \pgfmathsetmacro{\amp}{0.6 * \utterance[\i]}
            \pgfmathparse{int(\i==0||\i==1||\i==4||\i==8||\i==9)}
            \ifnum\pgfmathresult=1
                \draw[color=#2] (\xx pt, \y - \amp pt) -- (\xx pt, \y + \amp pt);
            \else
                \draw[color=#3] (\xx pt, \y - \amp pt) -- (\xx pt, \y + \amp pt);
            \fi
        }
    \end{scope}
}
\newcommand{\drawConversationWav}[3]{
    \path (#1); 
    \pgfgetlastxy{\startX}{\y}
    \begin{scope}[local bounding box=#3]
        \pgfmathsetmacro{\linespacing}{1.0}
        \foreach \i in {0,...,32} {
            \pgfmathsetmacro{\xx}{\startX + \i * \linespacing}
            \pgfmathsetmacro{\amp}{0.6 * \conversation[\i]}%
            \draw[color=#2] (\xx pt, \y - \amp pt) -- (\xx pt, \y + \amp pt);
        }
    \end{scope}
}

\newcommand{\coloredVector}[2]{%
    \foreach \y in {1,...,#1} {
        \pgfmathtruncatemacro\colorindex{#2[\y-1]}
        \fill[mycolor\colorindex] (0em,\y*0.25em-0.25em) rectangle (0.25em,\y*0.25em);
    }
}

\definecolor{mycolor1}{HTML}{FFD5E5}
\definecolor{mycolor2}{HTML}{FFAACC}
\definecolor{mycolor3}{HTML}{FF8082}
\definecolor{mycolor4}{HTML}{FF5599}
\definecolor{mycolor5}{HTML}{FF2A7F}
\definecolor{mycolor6}{HTML}{FF0066}
\definecolor{mycolor7}{HTML}{D40055}
\definecolor{mycolor8}{HTML}{AA0044}
\definecolor{mycolor9}{HTML}{D7EEF4}
\definecolor{mycolor10}{HTML}{AFDDE9}
\definecolor{mycolor11}{HTML}{87CDDE}
\definecolor{mycolor12}{HTML}{5FBCD3}
\definecolor{mycolor13}{HTML}{37ABC6}
\definecolor{mycolor14}{HTML}{2C89A0}
\definecolor{mycolor15}{HTML}{216776}
\definecolor{mycolor16}{HTML}{164450}

\def\embedfirst{{%
13,15,11,12,10,14,9,13,
}}
\def\embedsecond{{%
8,3,6,2,3,6,7,3,
}}


\begin{scope}[local bounding box=wavs]
    \drawSingleWav{$(0pt,0pt)$}{spkblue}{wavs_wav1};
\end{scope}
\node[block,draw=none,text width=6em,font=\bfseries\footnotesize,above=2em of wavs] {Pretraining\\(identification)};

\node[block,draw=none,text width=6em,font=\scriptsize,above=0em of wavs] {1-spk};
\node[block_frozen,below=0.5em of wavs] (feat) {Log mel Fbank};
\node[block,below=0.5em of feat] (encoder) {Encoder (SID)};
\node[block,below=0.5em of encoder] (pool) {Pool\,+\,Linear};

\node[rectangle,draw,text width=1*0.25em,
        text height=8*0.25em,inner sep=0,below=6.35em+31.6pt of $(wavs_wav1.east)!0.5!(wavs_wav1.west)$,
        path picture={
            \coloredVector{8}{\embedfirst};
            \draw[step=0.25em, black] (path picture bounding box.south west) grid (path picture bounding box.north east);
        }
    ] (spk1_emb) {};
\node[spkblue,inner sep=0,font=\scriptsize,below=0.2em of spk1_emb] {Spk 1};

\drawConversationWav{$(wavs) + (5.5em,0em)$}{black}{wav2};
\node[block,draw=none,text width=6em,font=\bfseries\footnotesize,above=2em of wav2] {Pretraining\\(diarization)};
\node[block,draw=none,text width=6em,font=\scriptsize,above=0em of wav2] {Simulated or\\Real (compound)};
\node[block_frozen,below=0.5em of wav2] (feat2) {Log mel Fbank};
\node[block_frozen,below=0.5em of feat2] (encoder2) {Encoder (SID)};
\node[block_frozen,below=0.5em of encoder2] (pool2) {Pool\,+\,Linear};
\node[block,below=0.5em of pool2] (encoder_diar2) {Encoder (DIA)};
\node[block,below=0.5em of encoder_diar2] (diar2) {DIA backend};
\node[anchor=north west,spkblue,text width=2em,inner sep=0,font=\scriptsize] at ($(diar2.south west) + (0,-0.7em)$) (spk1_label) {Spk 1};
\draw[color=spkblue] ($(spk1_label.base east) + (3pt,0pt)$) -- ++(3pt,0pt) -- ++(0pt,3pt) -- ++(5pt,0pt) -- ++(0pt,-3pt) -- ++(10pt,0pt) -- ++(0pt,3pt) -- ++(7pt,0pt) -- ++(0pt,-3pt) -- ++(7pt,0pt);
\node[anchor=north west,spkred,text width=2em,inner sep=0,font=\scriptsize] at ($(diar2.south west) + (0,-1.5em)$) (spk2_label) {Spk 2};
\draw[color=spkred] ($(spk2_label.base east) + (3pt,0pt)$) -- ++(14pt,0pt) -- ++(0pt,3pt) -- ++(8pt,0pt) -- ++(0pt,-3pt) -- ++(10pt,0pt);
\node[anchor=north west,spkgreen,text width=2em,inner sep=0,font=\scriptsize] at ($(diar2.south west) + (0,-2.3em)$) (spk3_label) {Spk 3};
\draw[color=spkgreen] ($(spk3_label.base east) + (3pt,0pt)$) -- ++(1pt,0pt) -- ++(0pt,3pt) -- ++(3pt,0pt) -- ++(0pt,-3pt) -- ++(7pt,0pt) -- ++(0pt,3pt) -- ++(5pt,0pt) -- ++(0pt,-3pt) -- ++(16pt,0pt);
\node[anchor=north west,spkdarkblue,text width=2em,inner sep=0,font=\scriptsize] at ($(diar2.south west) + (0,-3.1em)$) (spk4_label) {Spk 4};
\draw[color=spkdarkblue] ($(spk4_label.base east) + (3pt,0pt)$) -- ++(28pt,0pt) -- ++(0pt,3pt) -- ++(4pt,0pt);

\drawConversationWav{$(wav2) + (5.5em,0em)$}{black}{wav3};
\node[block,draw=none,text width=6em,font=\bfseries\footnotesize,above=2em of wav3] {Finetuning\\(diarization)};
\node[block,draw=none,text width=5.5em,font=\scriptsize,above=0em of wav3] {Real\\(specific domain)};
\node[block_frozen,below=0.5em of wav3] (feat3) {Log mel Fbank};
\node[block_frozen,below=0.5em of feat3] (encoder3) {Encoder (SID)};
\node[block_frozen,below=0.5em of encoder3] (pool3) {Pool\,+\,Linear};
\node[block,below=0.5em of pool3] (encoder_diar3) {Encoder (DIA)};
\node[block,below=0.5em of encoder_diar3] (diar3) {DIA backend};
\node[anchor=north west,spkblue,text width=2em,inner sep=0,font=\scriptsize] at ($(diar3.south west) + (0,-0.7em)$) (spk1_label) {Spk 1};
\draw[color=spkblue] ($(spk1_label.base east) + (3pt,0pt)$) -- ++(3pt,0pt) -- ++(0pt,3pt) -- ++(5pt,0pt) -- ++(0pt,-3pt) -- ++(10pt,0pt) -- ++(0pt,3pt) -- ++(7pt,0pt) -- ++(0pt,-3pt) -- ++(7pt,0pt);
\node[anchor=north west,spkred,text width=2em,inner sep=0,font=\scriptsize] at ($(diar3.south west) + (0,-1.5em)$) (spk2_label) {Spk 2};
\draw[color=spkred] ($(spk2_label.base east) + (3pt,0pt)$) -- ++(14pt,0pt) -- ++(0pt,3pt) -- ++(8pt,0pt) -- ++(0pt,-3pt) -- ++(10pt,0pt);
\node[anchor=north west,spkgreen,text width=2em,inner sep=0,font=\scriptsize] at ($(diar3.south west) + (0,-2.3em)$) (spk3_label) {Spk 3};
\draw[color=spkgreen] ($(spk3_label.base east) + (3pt,0pt)$) -- ++(1pt,0pt) -- ++(0pt,3pt) -- ++(3pt,0pt) -- ++(0pt,-3pt) -- ++(7pt,0pt) -- ++(0pt,3pt) -- ++(5pt,0pt) -- ++(0pt,-3pt) -- ++(16pt,0pt);
\node[anchor=north west,spkdarkblue,text width=2em,inner sep=0,font=\scriptsize] at ($(diar3.south west) + (0,-3.1em)$) (spk4_label) {Spk 4};
\draw[color=spkdarkblue] ($(spk4_label.base east) + (3pt,0pt)$) -- ++(28pt,0pt) -- ++(0pt,3pt) -- ++(4pt,0pt);
\draw[->] (wavs) -- (feat);
\draw[->] (feat) -- (encoder);
\draw[->] (encoder) -- (pool);
\draw[->] (pool) -- ($(pool.south) + (0,-0.5em-31.6pt)$);

\draw[->] (wav2) -- (feat2);
\draw[->] (feat2) -- (encoder2);
\draw[->] (encoder2) -- (pool2);
\draw[->] (pool2) -- (encoder_diar2);
\draw[->] (encoder_diar2) -- (diar2);
\draw[->] (diar2) -- ($(diar2.south) + (0,-0.5em)$);

\draw[->] (wav3) -- (feat3);
\draw[->] (feat3) -- (encoder3);
\draw[->] (encoder3) -- (pool3);
\draw[->] (pool3) -- (encoder_diar3);
\draw[->] (encoder_diar3) -- (diar3);
\draw[->] (diar3) -- ($(diar3.south) + (0,-0.5em)$);

\draw[->,very thick,decorate,decoration={snake,amplitude=0.05em,segment length=0.5em}] (pool) -- (pool2);
\draw[->,very thick,decorate,decoration={snake,amplitude=0.05em,segment length=0.5em}] (pool2) -- (pool3);
\draw[->,very thick,decorate,decoration={snake,amplitude=0.05em,segment length=0.5em}] (encoder) -- (encoder2);
\draw[->,very thick,decorate,decoration={snake,amplitude=0.05em,segment length=0.5em}] (encoder2) -- (encoder3);
\draw[->,very thick,decorate,decoration={snake,amplitude=0.05em,segment length=0.5em}] (encoder_diar2) -- (encoder_diar3);
\draw[->,very thick,decorate,decoration={snake,amplitude=0.05em,segment length=0.5em}] (diar2) -- (diar3);

\end{tikzpicture}\label{fig:conventional_embed}}
    \hfill
    \subfloat[Proposed multi-speaker SID pretraining]{\makebox[4.9cm][c]{\begin{tikzpicture}[semithick,auto,
block/.style={
    rectangle,
    draw,
    text centered,
    text width=5.6em,
    inner sep=2pt,
    minimum height=1em,
    minimum width=5.6em,
    font=\footnotesize},
block_frozen/.style={
    rectangle,
    draw,
    dotted,
    text centered,
    text width=5.6em,
    inner sep=2pt,
    minimum height=1em,
    minimum width=5.6em,
    font=\footnotesize},
label/.style={
    draw=none,
    align=center,
    font=\small,
    inner sep=0,
    outer sep=0
},
]

\definecolor{spkblue}{HTML}{0072BC}
\definecolor{spkred}{HTML}{DB3D23}
\definecolor{spkgreen}{HTML}{008770}
\definecolor{spkcyan}{HTML}{0FC8F2}
\definecolor{spkteal}{HTML}{2CD5B6}
\definecolor{spkdarkblue}{HTML}{001973}

\def\noise{{%
1,1,2,3,2,1,1,2,2,3,3,2,1,
}}
\def\utterance{{%
3,4,6,3,2,2,4,6,7,5,3,2,3,
}}
\def\conversation{{%
3,4,6,3,2,2,4,6,7,5,3,2,1,
2,1,2,4,5,4,3,5,7,6,4,2,
1,1,2,4,3,2,3,5,6,5,2,2,
}}

\newcommand{\drawNoiseWav}[3]{
    \path (#1); 
    \pgfgetlastxy{\startX}{\y}
    \begin{scope}[local bounding box=#3]
        \pgfmathsetmacro{\linespacing}{1.0}
        \foreach \i in {0,...,12} {
            \pgfmathsetmacro{\xx}{\startX + \i * \linespacing}
            \pgfmathsetmacro{\amp}{0.6 * \noise[\i]}%
            \draw[color=#2] (\xx pt, \y - \amp pt) -- (\xx pt, \y + \amp pt);
        }
    \end{scope}
}
\newcommand{\drawSingleWav}[3]{
    \path (#1); 
    \pgfgetlastxy{\startX}{\y}
    \begin{scope}[local bounding box=#3]
        \pgfmathsetmacro{\linespacing}{1.0}
        \foreach \i in {0,...,12} {
            \pgfmathsetmacro{\xx}{\startX + \i * \linespacing}
            \pgfmathsetmacro{\amp}{0.6 * \utterance[\i]}%
            \draw[color=#2] (\xx pt, \y - \amp pt) -- (\xx pt, \y + \amp pt);
        }
    \end{scope}
}
\newcommand{\drawMultiWav}[4]{    
    \path (#1); 
    \pgfgetlastxy{\startX}{\y}
    \begin{scope}[local bounding box=#4]
        \pgfmathsetmacro{\linespacing}{1.0}
        \foreach \i in {0,...,12} {
            \pgfmathsetmacro{\xx}{\startX + \i * \linespacing}
            \pgfmathsetmacro{\amp}{0.6 * \utterance[\i]}
            \pgfmathparse{int(\i==0||\i==1||\i==4||\i==8||\i==9)}
            \ifnum\pgfmathresult=1
                \draw[color=#2] (\xx pt, \y - \amp pt) -- (\xx pt, \y + \amp pt);
            \else
                \draw[color=#3] (\xx pt, \y - \amp pt) -- (\xx pt, \y + \amp pt);
            \fi
        }
    \end{scope}
}
\newcommand{\drawConversationWav}[3]{
    \path (#1); 
    \pgfgetlastxy{\startX}{\y}
    \begin{scope}[local bounding box=#3]
        \pgfmathsetmacro{\linespacing}{1.0}
        \foreach \i in {0,...,32} {
            \pgfmathsetmacro{\xx}{\startX + \i * \linespacing}
            \pgfmathsetmacro{\amp}{0.6 * \conversation[\i]}%
            \draw[color=#2] (\xx pt, \y - \amp pt) -- (\xx pt, \y + \amp pt);
        }
    \end{scope}
}

\newcommand{\coloredVector}[2]{%
    \foreach \y in {1,...,#1} {
        \pgfmathtruncatemacro\colorindex{#2[\y-1]}
        \fill[mycolor\colorindex] (0em,\y*0.25em-0.25em) rectangle (0.25em,\y*0.25em);
    }
}

\definecolor{mycolor1}{HTML}{FFD5E5}
\definecolor{mycolor2}{HTML}{FFAACC}
\definecolor{mycolor3}{HTML}{FF8082}
\definecolor{mycolor4}{HTML}{FF5599}
\definecolor{mycolor5}{HTML}{FF2A7F}
\definecolor{mycolor6}{HTML}{FF0066}
\definecolor{mycolor7}{HTML}{D40055}
\definecolor{mycolor8}{HTML}{AA0044}
\definecolor{mycolor9}{HTML}{D7EEF4}
\definecolor{mycolor10}{HTML}{AFDDE9}
\definecolor{mycolor11}{HTML}{87CDDE}
\definecolor{mycolor12}{HTML}{5FBCD3}
\definecolor{mycolor13}{HTML}{37ABC6}
\definecolor{mycolor14}{HTML}{2C89A0}
\definecolor{mycolor15}{HTML}{216776}
\definecolor{mycolor16}{HTML}{164450}

\def\embedfirst{{%
13,15,11,12,10,14,9,13,
}}
\def\embedsecond{{%
8,3,6,2,3,6,7,3,
}}


\begin{scope}[local bounding box=wavs]
    \drawNoiseWav{$(0pt,0pt)$}{gray}{wavs_noise};
    \drawSingleWav{$(20pt,0pt)$}{spkblue}{wavs_wav1};
    \drawMultiWav{$(40pt,0pt)$}{spkblue}{spkred}{wavs_wav2};
\end{scope}

\node[block,draw=none,text width=6.3em,outer sep=0,font=\scriptsize,above=0em of wavs] {Noise\,/\,1-spk\,/\,2-spk};
\node[block,draw=none,text width=6em,font=\bfseries\footnotesize,above=2em of wavs] {Pretraining\\(identification)};
\node[block_frozen,below=0.5em of wavs] (feat) {Log mel Fbank};
\node[block,below=0.5em of feat] (encoder) {Encoder};
\node[block,below=0.5em of encoder] (pool) {Pool\,+\,Linear};
\node[rectangle,draw,text width=1*0.25em,dotted,dash pattern=on 1pt off 1pt,
        text height=8*0.25em,inner sep=0,below=6.35em+31.6pt of $(wavs_noise.east)!0.5!(wavs_noise.west)$,
        path picture={
            \draw[step=0.25em, black, dotted,dash pattern=on 1pt off 1pt] (path picture bounding box.south west) grid (path picture bounding box.north east);
        }
    ] (noise_emb) {};
\node[gray,inner sep=0,font=\scriptsize,below=0.2em of noise_emb] {N/A};
\node[rectangle,draw,text width=1*0.25em,
        text height=8*0.25em,inner sep=0,below=6.35em+31.6pt of $(wavs_wav1.east)!0.5!(wavs_wav1.west)$,
        path picture={
            \coloredVector{8}{\embedfirst};
            \draw[step=0.25em, black] (path picture bounding box.south west) grid (path picture bounding box.north east);
        }
    ] (spk1_emb) {};
\node[spkblue,inner sep=0,font=\scriptsize,below=0.2em of spk1_emb] {Spk 1};
\node[rectangle,draw,text width=1*0.25em,
        text height=8*0.25em,inner sep=0,below left=6.35em+31.6pt and 0.1em of $(wavs_wav2.east)!0.5!(wavs_wav2.west)$,
        path picture={
            \coloredVector{8}{\embedfirst};
            \draw[step=0.25em, black] (path picture bounding box.south west) grid (path picture bounding box.north east);
        }
    ] (spk2_emb) {};
\node[rectangle,draw,text width=1*0.25em,
        text height=8*0.25em,inner sep=0,below right=6.35em+31.6pt and 0.1em of $(wavs_wav2.east)!0.5!(wavs_wav2.west)$,
        path picture={
            \coloredVector{8}{\embedsecond};
            \draw[step=0.25em, black] (path picture bounding box.south west) grid (path picture bounding box.north east);
        }
    ] (spk3_emb) {};
\node[spkblue,inner sep=0,font=\scriptsize,below left=0.2em and 0.28em of $(spk2_emb.south)!0.5!(spk3_emb.south)$] {1};
\node[black,inner sep=0,font=\scriptsize,below=0.2em of $(spk2_emb.south)!0.5!(spk3_emb.south)$] {\&};
\node[spkred,inner sep=0,font=\scriptsize,below right=0.2em and 0.28em of $(spk2_emb.south)!0.5!(spk3_emb.south)$] {2};

\drawConversationWav{$(wavs) + (5.5em,0em)$}{black}{wav2};
\node[block,draw=none,text width=6em,font=\bfseries\footnotesize,above=2em of wav2] {Finetuning\\(diarization)};
\node[block,draw=none,text width=5.5em,font=\scriptsize,above=0em of wav2] {Real\\(specific domain)};
\node[block_frozen,below=0.5em of wav2] (feat2) {Log mel Fbank};
\node[block,below=0.5em of feat2] (encoder2) {Encoder};
\node[block,below=0.5em of encoder2] (diar) {DIA backend};
\node[anchor=north west,spkblue,text width=2em,inner sep=0,font=\scriptsize] at ($(diar.south west) + (0,-0.7em-31.6pt)$) (spk1_label) {Spk 1};
\draw[color=spkblue] ($(spk1_label.base east) + (3pt,0pt)$) -- ++(3pt,0pt) -- ++(0pt,3pt) -- ++(5pt,0pt) -- ++(0pt,-3pt) -- ++(10pt,0pt) -- ++(0pt,3pt) -- ++(7pt,0pt) -- ++(0pt,-3pt) -- ++(7pt,0pt);
\node[anchor=north west,spkred,text width=2em,inner sep=0,font=\scriptsize] at ($(spk1_label.north west) + (0,-0.8em)$) (spk2_label) {Spk 2};
\draw[color=spkred] ($(spk2_label.base east) + (3pt,0pt)$) -- ++(14pt,0pt) -- ++(0pt,3pt) -- ++(8pt,0pt) -- ++(0pt,-3pt) -- ++(10pt,0pt);
\node[anchor=north west,spkgreen,text width=2em,inner sep=0,font=\scriptsize] at ($(spk2_label.north west) + (0,-0.8em)$) (spk3_label) {Spk 3};
\draw[color=spkgreen] ($(spk3_label.base east) + (3pt,0pt)$) -- ++(1pt,0pt) -- ++(0pt,3pt) -- ++(3pt,0pt) -- ++(0pt,-3pt) -- ++(7pt,0pt) -- ++(0pt,3pt) -- ++(5pt,0pt) -- ++(0pt,-3pt) -- ++(16pt,0pt);
\node[anchor=north west,spkdarkblue,text width=2em,inner sep=0,font=\scriptsize] at ($(spk3_label.north west) + (0,-0.8em)$) (spk4_label) {Spk 4};
\draw[color=spkdarkblue] ($(spk4_label.base east) + (3pt,0pt)$) -- ++(28pt,0pt) -- ++(0pt,3pt) -- ++(4pt,0pt);

\draw[->] (wavs) -- (feat);
\draw[->] (feat) -- (encoder);
\draw[->] (encoder) -- (pool);
\draw[->] (pool) -- ($(pool.south) + (0,-0.5em-31.6pt)$);

\draw[->] (wav2) -- (feat2);
\draw[->] (feat2) -- (encoder2);
\draw[->] (encoder2) -- (diar);
\draw[->] (diar) -- ($(diar.south) + (0,-0.5em-31.6pt)$);
\draw[->,very thick,decorate,decoration={snake,amplitude=0.05em,segment length=0.5em}] (encoder) -- (encoder2);

\end{tikzpicture}}\label{fig:proposed}}
    \caption{Comparison of pretraining strategies. SID: speaker identification, DIA: speaker diarization.}\label{fig:comparison}
\end{figure*}

\subsection{End-to-end neural diarization}
EEND is initially proposed as a single-modeled clustering-free diarization method \cite{fujita2019end1}.
It generates frame-wise posteriors of speech activity for each of $S$ speakers $[\vect{p}_t]_{t=1}^{T}\in\left(0,1\right)^{S\times T}$ from input frame-wise acoustic features $\left[\vect{x}_t\right]_{t=1}^T\in\mathbb{R}^{D\times T}$ using a neural network.
The earliest model consists of an encoder and a diarization backend, each of which can be written as follows:
\begin{align}
    \vect{e}_1,\dots,\vect{e}_T&=f_\text{enc}^\text{(DIA)}(\vect{x}_1,\dots,\vect{x}_T),\label{eq:enc_dia}\\
    \vect{p}_1,\dots,\vect{p}_T&=g_\text{diar}(\vect{e}_1,\dots,\vect{e}_T).\label{eq:backend_dia}
\end{align}
The encoder $f_\text{enc}^\text{(DIA)}$ transforms the frame-wise features into frame-wise embeddings $[\vect{e}_t]_{t=1}^T\in\mathbb{R}^{D\times T}$ and the diarization backend $g_\text{diar}$ further transforms them into frame-wise posterior probabilities of speech/non-speech for each speaker \([\vect{p}_t]_{t=1}^T\).

The network is trained to minimize binary cross entropy between the posteriors and the ground-truth activities $\left[\vect{y}_t\right]_{t=1}^T\in\left\{0,1\right\}^{S\times T}$.
The difficulty regarding the training is that the order of speakers should not affect the optimization.
EEND employs permutation-free loss based on binary cross-entropy~\cite{fujita2019end1} or powerset cross-entropy~\cite{plaquet2023powerset} to cope with this difficulty.

Early models use a position-wise feed-forward network with sigmoid activation for the diarization backend, i.e., $g_\text{diar}:\mathbb{R}^D\to(0,1)^{S}$, limiting EEND to at most $S$ speakers~\cite{fujita2019end1}.
To break this limitation, some methods introduced block-wise processing followed by clustering to integrate the block-wise results \cite{kinoshita2021integrating,horiguchi2021towards}.
Clustering determines the number of speakers dynamically, unconstrained by the network architecture.
This paper adopts this approach and focuses on improving block-level diarization performance while the clustering part is out of scope.

\subsection{Multi-speaker identification with recursive pooling}\label{sec:recursive_attentive_pooling}
The process inside common single-speaker embedding extractors~\cite{desplanques2020ecapatdnn,yakovlev2024reshape} can be described as follows:
\begin{align}
\vect{e}_1,\dots,\vect{e}_T&=f_\text{enc}^\text{(SID)}(\vect{x}_1,\dots,\vect{x}_T),\label{eq:enc_sid}\\
\vect{e}'&=g_\text{pool}(\vect{e}_1,\dots,\vect{e}_T)\label{eq:pool_sid},\\
\vect{v}&=h_\text{linear}(\vect{e}').\label{eq:fc_sid}
\end{align}
Each equation represents i) transformation of frame-wise acoustic features into frame-wise embeddings via the encoder $f_\text{enc}^\text{(SID)}$ in \autoref{eq:enc_sid}, ii) aggregation into a single embedding $\vect{e}'$ via $g_\text{pool}$ in \autoref{eq:pool_sid}, and iii) dimensionality reduction with a linear layer $h_\text{linear}$ to compute a speaker embedding $\vect{v}$ in \autoref{eq:fc_sid}.

A recent framework~\cite{horiguchi2024recursive} extends this to extract embeddings $\vect{e}'$ for multiple speakers by replacing \autoref{eq:pool_sid} with
\begin{equation}
\vect{e}'_1,\dots,\vect{e}'_S=g_\text{pool}(\vect{e}_1,\dots,\vect{e}_T).
\end{equation}
Each of $\{\vect{e}'_s\}_{s=1}^S$ is then used to compute a speaker embedding $\{\vect{v}_s\}_{s=1}^S$ via \autoref{eq:fc_sid}.
More specifically, attention weights for pooling are recursively calculated to extract multiple embeddings.
These weights also determine when to stop recursive inference, thus estimating the number of speakers.
The network is trained to minimize speaker identification loss using $\vect{v}_s$ and speaker counting loss, so the encoder learns to separate speakers internally, fulfilling the requirement of speaker diarization.

\section{Pretraining strategies of EEND models}
\subsection{Baseline 1: Diarization pretraining}
Most EEND methods rely on pretraining using large-scale conversational datasets with permutation-free diarization loss~\cite{fujita2019end1,kinoshita2021integrating,horiguchi2021towards,bredin2023pyannote,plaquet2023powerset,maiti2021endtoend,yamashita2022improving,landini2022from}, referred to here as DIA pretraining (\autoref{fig:conventional_eend}).
Pretraining and finetuning share the same model architecture: an encoder (\(f_\text{enc}^\text{(DIA)}\) in \autoref{eq:enc_dia}) using, e.g., bi-directional long short-term memories (BLSTMs)~\cite{fujita2019end1} or Transformers~\cite{kinoshita2021advances,horiguchi2021towards}, and a lightweight backend (\(g_\text{diar}\) in \autoref{eq:backend_dia}).
The large-scale conversational dataset for pretraining can be obtained by either simulating conversations using single-speaker utterances or combining real datasets from various domains.
However, each approach presents challenges: simulated data is often low-quality and storage-intensive, while even compounded real datasets remain insufficient in amount.

\subsection{Baseline 2: Multi-stage pretraining of speaker identification and diarization}
Some studies explored speaker embedding extractors for EEND~\cite{cord2023frame,alvarez2024leveraging}, but they aim to replace the input hand-crafted features with speaker embeddings extracted using a pretrained extractor as shown in \autoref{fig:conventional_embed}.
In this case, the parameters of the speaker identification model are frozen, and a diarization encoder, as large as in \autoref{fig:conventional_eend}, is added on top.
To preserve temporal resolution, pooling is performed using a sliding window. 
Improved speaker discrimination of the input features may help mitigate training data scarcity.
However, the large diarization encoder increases the model parameters, and diarization pretraining is required to train it from scratch.
Also, an encoder trained only on single-speaker utterances may struggle with overlaps and silences, which are crucial for diarization.

We preliminarily examined that local pooling still degraded performance due to blurred temporal resolution, and the large diarization encoder did not improve performance when the speaker identification encoder was not frozen.
Therefore, we characterize this approach by the use of a single-speaker identification model and the multi-stage training of SID and DIA.

\subsection{Proposed multi-speaker identification pretraining}
The proposed SID pretraining is shown in \autoref{fig:proposed}.
First, the encoder is pretrained on a multi-speaker identification task with pooling and linear layers.
In the finetuning step for diarization, we simply reuse the pretrained speaker identification encoder for speaker diarization, i.e., \(f_\text{enc}^\text{SID}=f_\text{enc}^\text{DIA}\), avoiding doubling the number of parameters.
We remove the pooling and linear layers, insert instead a diarization backend consisting of a single LSTM followed by a linear layer, and fine-tune the whole model on real data from a specific domain.

To prepare for diarization, with a focus on handling both silence and multiple speakers, the pretraining was conducted using the method in \autoref{sec:recursive_attentive_pooling}, which leverages recursive attentive pooling to extract multiple speakers' embeddings.
In practice, it is often assumed in diarization studies that the maximum number of speakers speaking simultaneously is two (as seen in approaches like power-set loss~\cite{plaquet2023powerset} or overlap-handling post-processing for clustering-based methods~\cite{landini2020but,bullock2020overlap,horiguchi2021endtoend}).
Following this assumption, we used audio containing 0 to 2 speakers for the pretraining.\footnote{Note that it is possible to train the diarization model during finetuning to handle a larger number of speakers.}
For dynamic mixing, we avoid additional data loading compared to training a standard single-speaker identification model by reusing speech and noise signals within a minibatch.
For example, two-speaker mixtures are created by summing two single-speaker speech signals, and zero-speaker samples are reused from the noise signals applied for on-the-fly augmentation of \{1,2\}-speaker utterances.
In addition, the duration of input in this pretraining stage is relatively short ($\sim$\SI{3}{\second}), so we only consider fully overlapped mixtures for two-speaker cases.
Thus, there is no need to pay special care to make dialogue act patterns resemble real conversations, freeing from simulating quality-sensitive conversational data.

\section{Experimental setup}

\begin{table}[t]
\centering
\setlength{\tabcolsep}{4pt}
\caption{Datasets used in our experiments.}\label{tbl:dataset}
\scalebox{0.778}{%
\begin{threeparttable}
\begin{tabular}{@{\hspace{2mm}\hspace{9.5pt}}lllcS[table-format=4]S[table-format=2]S[table-format=2]r@{}}
\toprule
&&&&\multicolumn{3}{c}{Hours}&\multicolumn{1}{c@{}}{\multirow{2.92}{*}{\makecell{Disk\\usage}}}\\\cmidrule(l{\tabcolsep}){5-7}
&Name & Abbr.& \#Spk & {Train} & {Val} & {Test} \\\midrule
&\makebox[0pt][l]{\tikz[remember picture,baseline]{\node(voxceleb){};}}VoxCeleb 1\&2~\cite{nagrani2020voxceleb} & -- & 1 & 2720 & 174 & 11 & \SI{302}{\giga\byte}\\\cmidrule(l{\tabcolsep}){2-8}
&\makebox[0pt][l]{\tikz[remember picture,baseline]{\node(simorg){};}}SimOrg \cite{fujita2019end1} & -- & 1--4 & 2778 & 28 & {--} & \SI{301}{\giga\byte}\\
&\makebox[0pt][l]{\tikz[remember picture,baseline]{\node(simnatural){};}}SimNatural \cite{yamashita2022improving} & -- & 1--4 & 2778 & 28 & {--}& \SI{301}{\giga\byte}\\\midrule
&\makebox[0pt][l]{\tikz[remember picture,baseline]{\node(aishell){};}}AISHELL-4~\cite{fu2021aishell}$^\dag$ & AS-4 & 3--7 & 105 & 2 & 13 & \SI{13}{\giga\byte}\\
&AliMeeting~\cite{yu2022m2met} & Ali & 2--4 & 111 & 4 & 11 & \SI{14}{\giga\byte}\\
&AMI (first channel)~\cite{carletta2007unleashing} & AMI & 3--5 & 80 & 10 & 9 & \SI{11}{\giga\byte}\\
&MagicData-RAMC~\cite{yang2022open} & RAMC & 2--3 &  150 & 10 & 21 & \SI{19}{\giga\byte}\\
&MSDWild (few)~\cite{liu2022msdwild}$^\dag$ & MSD & 2--4 & 64 & 2 & 10 & \SI{8}{\giga\byte}\\
&\makebox[0pt][l]{\tikz[remember picture,baseline]{\node(voxconverse){};}}VoxConverse~\cite{chung2020spot}$^\dag$ & VC & 1--21 & 18 & 2 & 44 & \SI{7}{\giga\byte}\\\cmidrule(l){2-8}
&\makebox[0pt][l]{\tikz[remember picture,baseline]{\node(compound){};}}Compound & -- & 1--21 & 528 & 30 & 107 & \SI{71}{\giga\byte}\\\bottomrule
\end{tabular}
\begin{tablenotes}
  \small
  \item[\dag] Since there is no official train/val split, we used the custom split in \cite{plaquet2025mamba}.
\end{tablenotes}
\begin{tikzpicture}[overlay,remember picture]
    \coordinate (braceStart) at ($(aishell.north west)+(-1mm,0.25em)$);
    \coordinate (braceEnd) at ($(voxconverse.south west)+(-1mm,0.25em)$);
    \draw[decorate, decoration={brace, mirror, amplitude=5pt}] (braceStart) -- (braceEnd);
    \coordinate (braceTip) at ($ (braceStart)!0.5!(braceEnd) + (-5pt,0) $);
    \draw[->,thin] ($(voxceleb.west) + (-1mm,0.25em-0.4mm)$) -- ++(-5pt-1.2mm+0.8mm,0) |- ($(simorg.west) + (-1mm,0.25em)$);
    \draw[->,thin] ($(voxceleb.west) + (-1mm,0.25em+0.4mm)$) -- ++(-5pt-1.2mm,0) |- ($(simnatural.west) + (-1mm,0.25em)$) node[midway,xshift=-2mm,yshift=4.8mm,rotate=90,anchor=base]{Simulate};
    \draw[->,thin] ($(braceTip) + (-0.2mm,0)$) -- ++(-1mm,0) |- ($(compound) + (-2mm,0.25em)$) node[midway,xshift=-2mm,yshift=7.2mm,rotate=90,anchor=base]{Compound};
\end{tikzpicture}
\end{threeparttable}%
}
\end{table}
 
\subsection{Dataset}
\autoref{tbl:dataset} lists the datasets used in our experiments, all monaural with a \SI{16}{\kHz} sampling rate and \SI{16}{\bit} depth.
The VoxCeleb 1\&2 dataset was used for SID pretraining and mixture generation.
We used two simulation protocols.
The first follows original EEND training: concatenating utterances interleaved by silence to generate speaker-wise long-form audio, and then summing them into single audio~\cite{fujita2019end1}.
We generated 50\,000 50-second mixtures per \{1,2,3,4\} speakers for training, and 500 for validation.
This yielded 2778 hours (10M seconds) of training portion, which is approximately the same size as the source dataset requiring about \SI{300}{\giga\byte} of storage.
The second method aligns utterances to form natural dialogue act patterns~\cite{yamashita2022improving}.
We refer to the datasets generated using the protocols above as SimOrg and SimNatural, respectively.
The voice activity detector in FunASR~\cite{gao2023funasr} was applied prior to mixture generation to remove as much silence as possible from source utterances.

As the real conversational datasets, we used the six datasets: AISHELL-4~\cite{fu2021aishell}, AliMeeting~\cite{yu2022m2met}, the first channel of array microphones in AMI~\cite{carletta2007unleashing}, MagicData-RAMC~\cite{yang2022open}, the few-talker set of MSDWild~\cite{liu2022msdwild}, and VoxConverse~\cite{chung2020spot}.
Also, the compound of the six real datasets is used in the experiments.

\subsection{Pretraining details}
In the main experiments, we used ECAPA-TDNN~\cite{desplanques2020ecapatdnn} and ReDimNet-B2~\cite{yakovlev2024reshape} as the SID/DIA encoders.
Each takes 80- and 72-dimensional log mel filterbank features extracted with \SI{25}{\ms} width and \SI{10}{\ms} shift as input, respectively.

In the DIA pretraining in \autoref{fig:conventional_eend}, the encoder is followed by a lightweight diarization backend consisting of a single BLSM and linear layer.
We trained for 30 epochs using the Adam optimizer~\cite{kingma2015adam}, linearly warming up the learning rate to 0.001 over 1\,000 iterations, then decaying it by 0.8 per epoch.

In SID pretraining, the encoder is followed by channel- and context-dependent attentive statistics pooling~\cite{desplanques2020ecapatdnn} and a linear layer to generate a 192-dimensional speaker embedding.
For the baseline single-speaker SID pretraining, as the simplified version of \autoref{fig:conventional_embed}, the minibatch size was set to 256.
For the proposed multi-speaker SID pretraining in \autoref{fig:proposed}, each minibatch consists of i) 256 single-speaker utterances, ii) 128 two-speaker mixtures, and iii) at most 128 zero-speaker (noise only) audios.
All the samples were cropped to 3 seconds long.
The noises were reused from those used for on-the-fly data augmentation applied at a probability of 0.5.
The other training details followed the protocol described in \cite{horiguchi2024recursive}.

We also examined the effect of multi-stage pretraining used in \autoref{fig:conventional_embed}.
We apply DIA pretraining with Compound data as the second stage, not only for the models pretrained via SID but also for those pretrained via DIA using the simulated datasets.
The training strategy follows the one used in the first stage.

Once the pretraining was completed, finetuning was conducted for each dataset with the same learning rate scheduling but with a peak learning rate of 0.0001.
For SID-pretraining, the same diarization backend was added after the encoder.

In all diarization training, mixtures were chunked into 10-second segments and a batch size was set to 32.
We used power-set loss \cite{plaquet2023powerset} with up to four speakers and at most two overlapping at a time, resulting in an 11-class classification problem.

\subsection{Evaluation}
As in pyannote 3.1~\cite{plaquet2023powerset}, local diarization used 10-second windows with 1-second shift.
We used the model averaged over the three best DER epochs on the validation set.
Diarization error rate (DER) without collar forgiveness was used as the metric.

\section{Results}
\subsection{Preliminary results of speaker verification}
\begin{table}
\caption{EERs (\%) on single- and multi-speaker verification.}
\label{tbl:sv_results}
\centering
\scalebox{0.778}{%
\begin{tabular}{@{}lS[table-format=2.1,table-space-text-post={M}]S[table-format=1.2]S[table-format=2.2]S[table-format=2.2]@{}}
\toprule
Encoder & \#Params & \textit{s vs. s} & \textit{s vs. m} & \textit{m vs. m}\\\midrule
ECAPA-TDNN (1-spk) & 14.7M & 0.88 & 24.51 & 35.26\\
ECAPA-TDNN (\{0,1,2\}-spk) & 14.9M & 1.22 & 6.40 & 12.03\\\midrule
ReDimNet-B2 (1-spk) & 5.1M & 0.69 & 26.83 & 36.91\\
ReDimNet-B2 (\{0,1,2\}-spk) & 5.2M & 0.99 & 5.15 & 10.11\\
\bottomrule
\end{tabular}%
}
\end{table}

We first report speaker verification results under three conditions: whether two single-speaker recordings are of the same speaker (\textit{s vs. s}), whether a two-speaker recording has the speaker in a single-speaker recording (\textit{s vs. s}), and whether two two-speaker recordings have the same speaker (\textit{m vs. m}).
For \textit{s vs. s}, we used the standard VoxCeleb 1-O set for evaluation, and for \textit{s vs. m} and \textit{m vs. m}, the extended versions of VoxCeleb 1-O used in the previous study were employed~\cite{horiguchi2024recursive}.

The results are shown in \autoref{tbl:sv_results}.
In either architecture, training with audio containing a variable number of speakers significantly improved the equal error rate (EER) in both \textit{s vs. m} and \textit{m vs. m}.
In seen conditions (i.e., 1-spk on \svss{} and \{0,1,2\}-spk on all the conditions), ReDimNet-B2 consistently outperformed ECAPA-TDNN, showing better results than those reported in ~\cite{horiguchi2024recursive}.
Notably, when two-speaker audio was not used during training, ReDimNet-B2's strength in \textit{s vs. s} did not extend to \textit{s vs. m} or \textit{m vs. m}.
This underscores the need for multi-speaker training to handle multiple speakers.

\subsection{Comparison of pretraining strategies}
\begin{table}
\centering
\sisetup{detect-weight,mode=text}
\renewrobustcmd{\bfseries}{\fontseries{b}\selectfont}
\renewrobustcmd{\boldmath}{}
\newrobustcmd{\B}{\bfseries}
\def\Uline#1{#1\llap{\uline{\phantom{#1}}}}
\caption{DERs (\%) with various pretraining strategies. To compare only the performance of local diarization, the clustering was performed in an oracle manner. The best and second best scores are \textbf{bolded} and \underline{underlined}, respectively.}\label{tbl:main_results}
\setlength{\tabcolsep}{4pt}
\subfloat[Encoder: ECAPA-TDNN]{%
\scalebox{0.778}{%
\begin{tabular}{@{}l@{\hspace{2pt}}l*{3}{S[table-format=2.2]}S[table-format=1.2]*{3}{S[table-format=2.2]}@{}}
\toprule
 && \multicolumn{6}{c}{Finetuning \& evaluation dataset} & \multicolumn{1}{c@{}}{\multirow{2.92}{*}{\makecell{Macro\\Avg.}}}\\\cmidrule(l{\tabcolsep}r{\tabcolsep}){3-8}
ID & Pretraining & {AS-4} & {Ali} & {AMI} & {RAMC} & {MSD} & {VC}\\\midrule
\texttt{a1} & None & 11.70 & 19.11 & 20.51 & 9.61 & 21.14 & 10.95 & 15.05\\
\texttt{a2} & DIA (Compound) & 10.34 & 17.84 & 19.50 & 9.33 & 20.46 & 9.63 & 14.52\\
\texttt{a3} & DIA (SimOrg) & 10.30 & 17.47 & 18.99 & 9.27 & 19.47 & 8.95 & 14.08\\
\texttt{a4} & DIA (SimNatural) & \Uline{9.89} & 17.07 & \B 17.90 & 8.89 & 19.35 & 8.61 & 13.62\\
\texttt{a5} & SID (1-spk) & 10.58 & \Uline{16.94} & 18.04 & \B 8.38 & \B 18.01 & \Uline{8.66} & \Uline{13.44}\\
\texttt{a6} & SID (\{0,1,2\}-spk) & \B 9.73 & \B 16.69 & \Uline{17.95} & \Uline{8.69} & \Uline{18.43} & \B 8.49 & \B 13.33\\\customdashline{1-9}
\multicolumn{9}{@{}l}{\textbf{+ 2nd-stage pretraining via DIA (Compound)}}\\
\texttt{a3'}& DIA (SimOrg) & \Uline{9.92} & 16.94 & 18.37 & 9.07 & 19.60 & 9.03 & 13.82 \\
\texttt{a4'} & DIA (SimNatural) & 10.17 & 17.02 & 18.86 & 9.01 & 19.26 & 8.86 & 13.86\\
\texttt{a5'} & SID (1-spk) & 10.26 & \B 15.92 & \Uline{17.31} & \B 8.52 & \B 17.52 & \B 8.27 & \B 12.97 \\
\texttt{a6'} & SID (\{0,1,2\}-spk) & \B 9.58 & \Uline{16.22} & \B 17.19 & \Uline{8.79} & \Uline{17.62} & \Uline{8.46} & \Uline{12.98}\\
\bottomrule
\end{tabular}%
}%
}\\
\subfloat[Encoder: ReDimNet-B2]{%
\scalebox{0.778}{%
\begin{tabular}{@{}l@{\hspace{2pt}}l*{3}{S[table-format=2.2]}S[table-format=1.2]*{3}{S[table-format=2.2]}@{}}
\toprule
&& \multicolumn{6}{c}{Finetuning \& evaluation dataset} & \multicolumn{1}{c@{}}{\multirow{2.92}{*}{\makecell{Macro\\Avg.}}}\\\cmidrule(l{\tabcolsep}r{\tabcolsep}){3-8}
ID & Pretraining & {AS-4} & {Ali} & {AMI} & {RAMC} & {MSD} & {VC}\\\midrule
\texttt{b1} & None & 11.65 & 18.00 & 19.19 & 8.96 & 20.20 & 10.51 & 14.75\\
\texttt{b2} & DIA (Compound) & 9.80 & 16.13 & 17.47 & \Uline{8.47} & 18.43 & 8.70 & 13.17\\
\texttt{b3} & DIA (SimOrg) & \Uline{9.58} & \Uline{15.67} & 17.10 & 8.65 & 17.47 & 8.05 & \Uline{12.75}\\
\texttt{b4} & DIA (SimNatural) & 10.07 & 15.70 & 17.24 & 8.51 & 17.48 & \B 7.81 & 12.80\\
\texttt{b5} & SID (1-spk) & 9.64 & 15.80 & \Uline{16.60} & 9.11 & \Uline{16.96} & 8.63 & 12.79\\
\texttt{b6} & SID (\{0,1,2\}-spk) & \B 9.23 & \B 15.05 & \B 16.44 & \B 8.04 & \B 15.52 & \Uline{8.03} & \B 12.05\\\customdashline{1-9}
\multicolumn{9}{@{}l}{\textbf{+ 2nd-stage pretraining via DIA (Compound)}}\\
\texttt{b3'} & DIA (SimOrg) & 9.58 & 15.67 & 17.10 & 8.65 & 17.47 & 8.05 & 12.75\\
\texttt{b4'} &DIA (SimNatural) & 9.67 & 15.82 & 17.14 & 8.42 & 17.89 & 8.68 & 12.94\\
\texttt{b5'} &SID (1-spk) & \Uline{8.95} & \Uline{14.49} & \Uline{15.53} & \B 8.09 & \Uline{15.97} & \Uline{7.88} & \Uline{11.82}\\
\texttt{b6'} &SID (\{0,1,2\}-spk) & \B 8.53 & \B 13.96 & \B 15.05 & \Uline{8.25} & \B 15.21 & \B 7.34 & \B 11.39\\
\bottomrule
\end{tabular}%
}%
}
\end{table}
To show the effectiveness of the proposed method, the following pretraining strategies are compared: no pretraining, DIA pretraining using Compound/SimOrg/SimNatural, and SID pretraining using single-speaker utterances and \{0,1,2\}-speaker utterances.
We used ECAPA-TDNN and ReDimNet-B2 for both speaker identification and diarization encoders.
To focus on local diarization performance, speaker assignment to global labels from local results was oracle-based here.

The results are shown in \autoref{tbl:main_results}.
We first confirmed that the pretraining using a simulated dataset is important even when compound data is available (\texttt{a1}--\texttt{a4} and \texttt{b1}--\texttt{b4}).
However, the impact of the simulation method varied by architecture.
Unlike standard EEND, the 10-second input might limit the importance of the simulation protocol, as it is too short to model dialogue act patterns.
The SID pretraining achieved DER comparable to or better than the DIA pretraining when based on 1-spk and performed even better with \{0,1,2\}-spk.
We can conclude that SID pretraining outperformed conventional DIA pretraining, and incorporating multi-speaker identification further enhanced the model’s suitability for diarization.

We also showed the results with the second-stage pretraining using the compound dataset.
It is noteworthy that the models pretrained with SID largely benefit from this additional pretraining, while those using DIA pretraining did not.
This is likely due to the acquisition of diarization-related capabilities, such as handling more speakers and partial overlaps.
The models initially pretrained using SID had room for improvement in this aspect, whereas those pretrained in a DIA manner using simulated data did not.
Moreover, the second pretraining stage helped acquire multi-speaker handling ability, reducing the gap between SID pretraining using 1-spk and \{0,1,2\}-spk.
For ECAPA-TDNN, this eliminated the advantages of \{0,1,2\}-speaker pretraining (\texttt{a5'} vs. \texttt{a6'}), while for ReDimNet, \{0,1,2\}-spk pretraining remained superior (\texttt{b5'} vs. \texttt{b6'}).

\subsection{Comparison to other baseline methods}
\begin{table}
\caption{DERs (\%) of baselines and the proposed method.}\label{tbl:results_comparison}
\renewrobustcmd{\bfseries}{\fontseries{b}\selectfont}
\renewrobustcmd{\boldmath}{}
\newrobustcmd{\B}{\bfseries}
\def\Uline#1{#1\llap{\uline{\phantom{#1}}}}
\setlength{\tabcolsep}{3pt}
\resizebox{\linewidth}{!}{%
\begin{tabular}{@{}l@{\hspace{1pt}}S[table-format=2.1,table-space-text-post={M}]*{3}{S[table-format=2.2]}S[table-format=1.2]S[table-format=2.2]S[table-format=1.2]S[table-format=2.2]@{}}
\toprule
& & \multicolumn{6}{c}{Finetuning \& evaluation dataset} & \multicolumn{1}{c@{}}{\multirow{2.92}{*}{\makecell{Macro\\Avg.}}}\\\cmidrule(l{\tabcolsep}r{\tabcolsep}){3-8}
Architecture & {\#Params} & {AS-4} & {Ali} & {AMI} & {RAMC} & {MSD} & {VC}\\\midrule
SincNet-BLSTM (DIA) & 1.5M & 12.55 & 21.86 & 22.96 & 14.57 & 27.16 & 11.81 & 18.49\\
WavLM-BLSTM (DIA) & 96.5M & 11.92 & 18.81 & 19.21 & 11.92 & 22.39 & 9.55 & 15.63\\\midrule
\texttt{a6}\ \ ECAPA-TDNN (SID) & 15.0M & 11.59 & 20.08 & 22.54 & 14.10 & 25.13 & 10.28 & 17.29\\
\texttt{a6'}\ \ + DIA (Compound) &  & 11.40 & 19.92 & 20.22 & 12.55 & 24.03 & 10.41 & 16.42\\
\texttt{b6}\ \ ReDimNet-B2 (SID) & 5.4M & 11.31 & 20.44 & 20.27 & \B 11.65 & 21.80 & \B 9.51 & 15.83\\
\texttt{b6'}\ \ + DIA (Compound) &  & \B 10.26 & \B 17.54 & \B 18.96 & 12.55 & \B 21.77 & 10.00 & \B 15.18\\
\bottomrule
\end{tabular}%
}
\end{table}

The evaluation in \autoref{tbl:main_results} is based on the less common condition of using encoders proposed for speaker embedding extraction in the context of diarization; thus, readers may wonder how this compares to common architectures used in diarization.
This subsection then compares the proposed method with the architectures commonly used in diarization studies.
One is the architecture used in pyannote.audio 3.1~\cite{bredin2023pyannote}, consists of SincNet~\cite{ravanelli2018speaker}, four-stacked BLSTMs, and two linear layers.
The other one leverages the encoder trained using large-scale datasets with SSL.
We used WavLM Base+~\cite{chen2022wavlm}, which is widely adopted for feature extractor of EEND~\cite{baroudi2023pyannote,tawara2024ntt,han2025leveraging}.
The parameters of WavLM were frozen from the pretrained weights.
The weighted sum of the outputs from all the intermediate layers was fed to the diarization backend consisting of four-stacked BLSTMs and two linear layers.
We used DIA pretraining with Compound data for those two methods.
Agglomerative hierarchical clustering with ResNet-based speaker embeddings implemented in pyannote.audio was used to integrate local diarization results.

The results are shown in \autoref{tbl:results_comparison}.
Our method achieved comparative performance to WavLM-BLSTM with ReDimNet-B2 with SID pretraining using \{0,1,2\}-speaker audio (\texttt{b6}) with 
Our method achieved performance comparable to WavLM-BLSTM with ReDimNet-B2 using SID pretraining on {0,1,2}-speaker audio (\texttt{b6}), while using only about \SI{6}{\percent} of the parameters.
The second-stage pretraining using the compound dataset brought additional performance improvement, outperforming WavLM-BLSTM (\texttt{b6}').
Considering that SSL models are trained over long durations on a lot of GPUs (e.g., 32~\cite{chen2022wavlm}) using a very large-scale dataset (e.g., 94k hours~\cite{chen2022wavlm}), these results underscore the potential for efficient pretraining in speaker-related tasks including diarization.

\section{Conclusion}
This paper demonstrated the effectiveness of multi-speaker SID pretraining for EEND.
The method is storage-friendly, simulation-agnostic, and outperformed diarization-based pretraining, with further gains from additional DIA pretraining.
Future work will include the method to perform local diarization and speaker embedding extraction in a single model.

\clearpage
\bibliographystyle{IEEEtran}
\bibliography{mybib}

\begin{thebibliography}{10}
\providecommand{\url}[1]{#1}
\csname url@samestyle\endcsname
\providecommand{\newblock}{\relax}
\providecommand{\bibinfo}[2]{#2}
\providecommand{\BIBentrySTDinterwordspacing}{\spaceskip=0pt\relax}
\providecommand{\BIBentryALTinterwordstretchfactor}{4}
\providecommand{\BIBentryALTinterwordspacing}{\spaceskip=\fontdimen2\font plus
\BIBentryALTinterwordstretchfactor\fontdimen3\font minus \fontdimen4\font\relax}
\providecommand{\BIBforeignlanguage}[2]{{%
\expandafter\ifx\csname l@#1\endcsname\relax
\typeout{** WARNING: IEEEtran.bst: No hyphenation pattern has been}%
\typeout{** loaded for the language `#1'. Using the pattern for}%
\typeout{** the default language instead.}%
\else
\language=\csname l@#1\endcsname
\fi
#2}}
\providecommand{\BIBdecl}{\relax}
\BIBdecl

\bibitem{boeddeker2018front}
C.~Boeddeker, J.~Heitkaemper, J.~Schmalenstoeer, L.~Drude, J.~Heymann, and R.~Haeb-Umbach, ``Front-end processing for the {CHiME-5} dinner party scenario,'' in \emph{Proc. CHiME-5}, 2018, pp. 35--40.

\bibitem{polok2024but}
A.~Polok, D.~Klement, J.~Han, {\v{S}}.~Sedl{\'a}{\v{c}}ek, B.~Yusuf, M.~Maciejewski, M.~S. Wiesner, and L.~Burget, ``{BUT/JHU} system description for {CHiME-8} {NOTSOFAR-1} challenge,'' in \emph{Proc. CHiME}, 2024, pp. 18--22.

\bibitem{polok2024dicow}
A.~Polok, D.~Klement, M.~Kocour, J.~Han, F.~Landini, B.~Yusuf, M.~Wiesner, S.~Khudanpur, J.~{\v{C}}ernock{\`y}, and L.~Burget, ``{DiCoW}: Diarization-conditioned whisper for target speaker automatic speech recognition,'' arXiv:2501.00114, 2024.

\bibitem{landini2022bayesian}
F.~Landini, J.~Profant, M.~Diez, and L.~Burget, ``Bayesian {HMM} clustering of x-vector sequences ({VBx}) in speaker diarization: Theory, implementation and analysis on standard tasks,'' \emph{Computer Speech \& Language}, vol.~71, p. 101254, 2022.

\bibitem{fujita2019end1}
Y.~Fujita, N.~Kanda, S.~Horiguchi, K.~Nagamatsu, and S.~Watanabe, ``End-to-end neural speaker diarization with permutation-free objectives,'' in \emph{Proc. Interspeech}, 2019, pp. 4300--4304.

\bibitem{kinoshita2021integrating}
K.~Kinoshita, M.~Delcroix, and N.~Tawara, ``Integrating end-to-end neural and clustering-based diarization: Getting the best of both worlds,'' in \emph{Proc. ICASSP}, 2021, pp. 7198--7202.

\bibitem{horiguchi2021towards}
S.~Horiguchi, S.~Watanabe, P.~Garc\'{i}a, Y.~Xue, Y.~Takashima, and Y.~Kawaguchi, ``Towards neural diarization for unlimited numbers of speakers using global and local attractors,'' in \emph{Proc. ASRU}, 2021, pp. 98--105.

\bibitem{bredin2023pyannote}
H.~Bredin, ``{pyannote.audio 2.1} speaker diarization pipeline: principle, benchmark, and recipe,'' in \emph{Proc. Interspeech}, 2023, pp. 1983--1987.

\bibitem{plaquet2023powerset}
A.~Plaquet and H.~Bredin, ``Powerset multi-class cross entropy loss for neural speaker diarization,'' in \emph{Proc. Interspeech}, 2023, pp. 3222--3226.

\bibitem{maiti2021endtoend}
S.~Maiti, H.~Erdogan, K.~Wilson, S.~Wisdom, S.~Watanabe, and J.~R. Hershey, ``End-to-end diarization for variable number of speakers with local-global networks and discriminative speaker embeddings,'' in \emph{Proc. ICASSP}, 2021, pp. 7183--7187.

\bibitem{yamashita2022improving}
N.~Yamashita, S.~Horiguchi, and T.~Homma, ``Improving the naturalness of simulated conversations for end-to-end neural diarization,'' in \emph{Proc. Odyssey}, 2022, pp. 133--140.

\bibitem{landini2022from}
F.~Landini, A.~Lozano-Diez, M.~Diez, and L.~Burget, ``From simulated mixtures to simulated conversations as training data for end-to-end neural diarization,'' in \emph{Proc. Interspeech}, 2022, pp. 5095--5099.

\bibitem{han2025leveraging}
J.~Han, F.~Landini, J.~Rohdin, A.~Silnova, M.~Diez, and L.~Burget, ``Leveraging self-supervised learning for speaker diarization,'' in \emph{Proc. ICASSP}, 2025.

\bibitem{horiguchi2024recursive}
S.~Horiguchi, A.~Ando, T.~Moriya, T.~Ashihara, H.~Sato, N.~Tawara, and M.~Delcroix, ``Recursive attentive pooling for extracting speaker embeddings from multi-speaker recordings,'' in \emph{Proc. SLT}, 2024, pp. 1219--1226.

\bibitem{cord2023frame}
T.~Cord-Landwehr, C.~Boeddeker, C.~Zoril{\u{a}}, R.~Doddipatla, and R.~Haeb-Umbach, ``Frame-wise and overlap-robust speaker embeddings for meeting diarization,'' in \emph{Proc. ICASSP}, 2023.

\bibitem{alvarez2024leveraging}
J.~I. Alvarez-Trejos, B.~Labrador, and A.~Lozano-Diez, ``Leveraging speaker embeddings in end-to-end neural diarization for two-speaker scenarios,'' in \emph{Proc. Odyssey}, 2024, pp. 107--114.

\bibitem{desplanques2020ecapatdnn}
B.~Desplanques, J.~Thienpondt, and K.~Demuynck, ``{ECAPA-TDNN}: Emphasized channel attention, propagation and aggregation in {TDNN} based speaker verification,'' in \emph{Proc. Interspeech}, 2020, pp. 3830--3834.

\bibitem{yakovlev2024reshape}
I.~Yakovlev, R.~Makarov, A.~Balykin, P.~Malov, A.~Okhotnikov, and N.~Torgashov, ``Reshape dimensions network for speaker recognition,'' in \emph{Proc. Interspeech}, 2024, pp. 3235--3239.

\bibitem{kinoshita2021advances}
K.~Kinoshita, M.~Delcroix, and N.~Tawara, ``Advances in integration of end-to-end neural and clustering-based diarization for real conversational speech,'' in \emph{Proc. Interspeech}, 2021, pp. 3565--3569.

\bibitem{landini2020but}
F.~Landini, S.~Wang, M.~Diez, L.~Burget, P.~Mat{\v{e}}jka, K.~{\v{Z}}mol{\'\i}kov{\'a}, L.~Mo{\v{s}}ner, A.~Silnova, O.~Plchot, O.~Novotn{\`y}, H.~Zeinali, and J.~Rohdin, ``{BUT} system for the {Second DIHARD Speech Diarization Challenge},'' in \emph{Proc. ICASSP}, 2020, pp. 6529--6533.

\bibitem{bullock2020overlap}
L.~Bullock, H.~Bredin, and L.~P. Garcia-Perera, ``Overlap-aware diarization: Resegmentation using neural end-to-end overlapped speech detection,'' in \emph{Proc. ICASSP}, 2020, pp. 7114--7118.

\bibitem{horiguchi2021endtoend}
S.~Horiguchi, P.~Garcia, Y.~Fujita, S.~Watanabe, and K.~Nagamatsu, ``End-to-end speaker diarization as post-processing,'' in \emph{Proc. ICASSP}, 2021, pp. 7188--7192.

\bibitem{nagrani2020voxceleb}
A.~Nagrani, J.~S. Chung, W.~Xie, and A.~Zisserman, ``{VoxCeleb}: Large-scale speaker verification in the wild,'' \emph{Computer Speech \& Language}, vol.~60, p. 101027, 2020.

\bibitem{fu2021aishell}
Y.~Fu, L.~Cheng, S.~Lv, Y.~Jv, Y.~Kong, Z.~Chen, Y.~Hu, L.~Xie, J.~Wu, H.~Bu, X.~Xu, J.~Du, and J.~Chen, ``{AISHELL-4}: An open source dataset for speech enhancement, separation, recognition and speaker diarization in conference scenario,'' in \emph{Proc. Interspeech}, 2021, pp. 3665--3669.

\bibitem{yu2022m2met}
F.~Yu, S.~Zhang, Y.~Fu, L.~Xie, S.~Zheng, Z.~Du, W.~Huang, P.~Guo, Z.~Yan, B.~Ma, X.~Xu, and H.~Bu, ``{M2MeT}: The {ICASSP} 2022 multi-channel multi-party meeting transcription challenge,'' in \emph{Proc. ICASSP}, 2022, pp. 6167--6171.

\bibitem{carletta2007unleashing}
J.~Carletta, ``Unleashing the killer corpus: experiences in creating the multi-everything {AMI Meeting Corpus},'' \emph{Language Resources and Evaluation}, vol.~41, no.~2, pp. 181--190, 2007.

\bibitem{yang2022open}
Z.~Yang, Y.~Chen, L.~Luo, R.~Yang, L.~Ye, G.~Cheng, J.~Xu, Y.~Jin, Q.~Zhang, P.~Zhang, L.~Xie, and Y.~Yan, ``Open source {MagicData-RAMC}: A rich annotated {Mandarin} conversational ({RAMC}) speech dataset,'' in \emph{Proc. Interspeech}, 2022, pp. 1736--1740.

\bibitem{liu2022msdwild}
T.~Liu, S.~Fan, X.~Xiang, H.~Song, S.~Lin, J.~Sun, T.~Han, S.~Chen, B.~Yao, S.~Liu, Y.~Wu, Y.~Qian, and K.~Yu, ``{MSDWild}: Multi-modal speaker diarization dataset in the wild,'' in \emph{Proc. Interspeech}, 2022, pp. 1476--1480.

\bibitem{chung2020spot}
J.~S. Chung, J.~Huh, A.~Nagrani, T.~Afouras, and A.~Zisserman, ``Spot the conversation: Speaker diarisation in the wild,'' in \emph{Proc. Interspeech}, 2020, pp. 299--303.

\bibitem{plaquet2025mamba}
A.~Plaquet, N.~Tawara, M.~Delcroix, S.~Horiguchi, A.~Ando, and S.~Araki, ``Mamba-based segmentation model for speaker diarization,'' in \emph{Proc. ICASSP}, 2025.

\bibitem{gao2023funasr}
Z.~Gao, Z.~Li, J.~Wang, H.~Luo, X.~Shi, M.~Chen, Y.~Li, L.~Zuo, Z.~Du, Z.~Xiao, and S.~Zhang, ``{FunASR}: A fundamental end-to-end speech recognition toolkit,'' in \emph{Proc. Interspeech}, 2023, pp. 1593--1597.

\bibitem{kingma2015adam}
D.~P. Kingma and J.~Ba, ``Adam: A method for stochastic optimization,'' in \emph{Proc. ICLR}, 2015.

\bibitem{ravanelli2018speaker}
M.~Ravanelli and Y.~Bengio, ``Speaker recognition from raw waveform with {SincNet},'' in \emph{Proc. SLT}, 2018, pp. 1021--1028.

\bibitem{chen2022wavlm}
S.~Chen, C.~Wang, Z.~Chen, Y.~Wu, S.~Liu, Z.~Chen, J.~Li, N.~Kanda, T.~Yoshioka, X.~Xiao \emph{et~al.}, ``{WavLM}: Large-scale self-supervised pre-training for full stack speech processing,'' \emph{IEEE Journal of Selected Topics in Signal Processing}, vol.~16, no.~6, pp. 1505--1518, 2022.

\bibitem{baroudi2023pyannote}
S.~Baroudi, H.~Bredin, A.~Plaquet, and T.~Pellegrini, ``{pyannote.audio} speaker diarization pipeline at {VoxSRC} 2023,'' The VoxCeleb Speaker Recognition Challenge, 2023.

\bibitem{tawara2024ntt}
N.~Tawara, M.~Delcroix, A.~Ando, and A.~Ogawa, ``{NTT} speaker diarization system for {CHiME-7}: Multi-domain, multi-microphone end-to-end and vector clustering diarization,'' in \emph{Proc. ICASSP}, 2024, pp. 11\,281--11\,285.

\end{thebibliography}

\end{document}